\documentclass[twocolumn,showpacs,preprintnumbers,amsmath,amssymb,10pt,a4paper,prl,aps]{revtex4-1} 

\usepackage{graphicx}% Include figure files
\usepackage{epsfig}
\usepackage[latin1]{inputenc}
\usepackage{amsmath}
\usepackage{amsfonts}
\usepackage{amssymb}
\usepackage{dcolumn}% Align table columns on decimal point
\usepackage{bm}% bold math

\begin{document}

\title{Control of the magnetic vortex chirality in Permalloy nanowires with asymmetric notches}%

\author{J. Brand\~{a}o}
\author{R. L. Novak}\altaffiliation{Current address: Universidade Federal de Santa Catarina, Campus Blumenau, Rua Pomerode 710, 89065-300, SC, Brazil}
\author{H. Lozano}
\author{P. R. Soledade}
\author{A. Mello}
\author{F. Garcia}
\author{L. C. Sampaio}

\affiliation{%
Centro Brasileiro de Pesquisas F\'{\i}sicas, Xavier Sigaud, 150, Rio de Janeiro, RJ, 22.290-180, Brazil
}%


\date{\today}% It is always \today, today,

\begin{abstract}

We have investigated the motion of vortex domain walls passing across non symmetric triangular notches in single Permalloy nanowires. We have measured hysteresis cycles using the focused magneto-optical Kerr effect before and beyond the notch, which allowed to probe  beyond the notch the occurrence probability of clockwise (CW) and counter-clockwise (CCW) walls in tail-to-tail (TT) and head-to-head (HH) configurations. We present experimental evidence of chirality flipping provided by the vortex -- notch interaction. With a low exit angle the probability of chirality flipping increases and here with the lowest angle of 15$^o$ the probability of propagation of the energetically favored domain wall configuration (CCW for TT or CW for HH walls) is $\approx 75\%$. Micromagnetic simulations reveal details of the chirality reversal dynamics. 

\end{abstract}


\maketitle

% beginning
\section{Introduction}

The manipulation and control of magnetization in small scales under presence of magnetic field, spin-polarized current, temperature gradient \cite{Jiang}, and electric field~\cite{Ohno} have been a subject of growing interest in the last decade, with important consequences to technological applications. One of the approaches is based on the control of the domain wall motion in nanowires, which can be used to build logical devices~\cite{Cowburn} and memories, such as the racetrack memory~\cite{Stuart}. Besides some parameters like the magnetization direction (in-plane or out-of-plane), wall type, and the external excitation, the presence of artificial defects like notches is very useful to control the wall position and movement. 

The vortex structure is observed in Nature in a wide range of phenomena, from tornadoes and hurricanes in atmospheric phenomena to superconductivity. In magnetism, it is present in confined structures like disks, triangles, squares and wires (or stripes) in micron and submicron scales. The magnetic vortex is characterized by the in-plane magnetization that circulates in the CW or CCW directions (or chiralities) and by the out-of-plane magnetization, at its core, that is polarized up or down~\cite{Chien}, leading to four independent combinations of chirality and polarity. 

The control and flipping of vortex core polarity have been extensively studied in the last years. It was demonstrated theoretically, experimentally and by micromagnetic simulations, that it is possible to switch and control the polarity by several ways, e.g., by the application of a magnetic field pulse~\cite{Pigeau}, rotating magnetic field, or spin polarized current. However, it is not straightforward to achieve a fine control of the vortex chirality. Probably, for this reason, so few papers have faced this issue~\cite{Cambel,Puig,Jaafar}. How to manipulate the chirality and polarity of vortex walls moving on magnetic stripes is a challenging issue, mainly because of the difficulty in obtaining good-to-noise ratio with single object measurements and not in arrays, as it has been done in most of the work reported so far.

Some recent papers have reported experimental~\cite{Bogart, Bogart2, Hayashi, Hayashi2, Eastwood, Meier}, and simulation~\cite{He} data about the dynamics of domain walls in Permalloy (Py) nanowires with notches, walls  driven by magnetic field and/or spin polarized current. These authors have demonstrated that notches with different geometries (squares or triangular) provide different potential barriers to the domain walls, which in turn can be transversal or vortex walls~\cite{Michael, Thiaville}. 

Although the dependence of the depinning field (or coercivity) with the wall chirality was already observed~\cite{Bogart, Jin, oommf}, to get more detailed information about the vortex wall -- notch interaction, measurements with an improved signal-to-noise ratio of focused MOKE is absolutely necessary. It allows to clearly define the shape of the hysteresis loop and determine precisely the coercivity of both chiralities, CW and CCW, and mainly their relative occurrence probability beyond the notch.

In this work, through the focused magneto-optical Kerr effect (MOKE) and micromagnetic simulations, we  investigate the dynamics of vortex walls traveling and interacting with triangular asymmetric notches on Permalloy (Py) nanowires. The main idea is varying the notch symmetry by changing the outgoing angle $\phi$ (as shown in Fig.\ref{Fig1}), to be able to continuously vary the potential barrier generated by the notch. Our main goal is to understand whether or not the vortex wall chirality can be tailored when the wall interacts with a defect with a particular symmetry, here, a notch, and what should be the vortex wall dynamics.

\section{Experiment}

Permalloy (Fe$_{80}$Ni$_{20}$) nanowires were fabricated by electron-beam lithography followed by sputtering deposition and lift-off. The Py films were grown onto Si(100)/SiO$_2$ using magnetron sputtering with a base pressure of $10 ^{-8}$ Torr. The width and thickness of the nanowires were chosen to favor walls as vortex domain walls. For the sake of presentation, all results shown herein are related to nanowires 435 nm wide, 30 nm thick and 100 $\mu$m long. Other wires with close dimensions and also exhibiting vortex domain walls showed similar behavior. Special care was taken in the lithography process to avoid defects between writing fields. Defects at the wire edges would provide an artificial increase in the coercivity, masking the role played by the notch.    

\begin{figure}[h] 
\includegraphics[width=0.9\columnwidth]{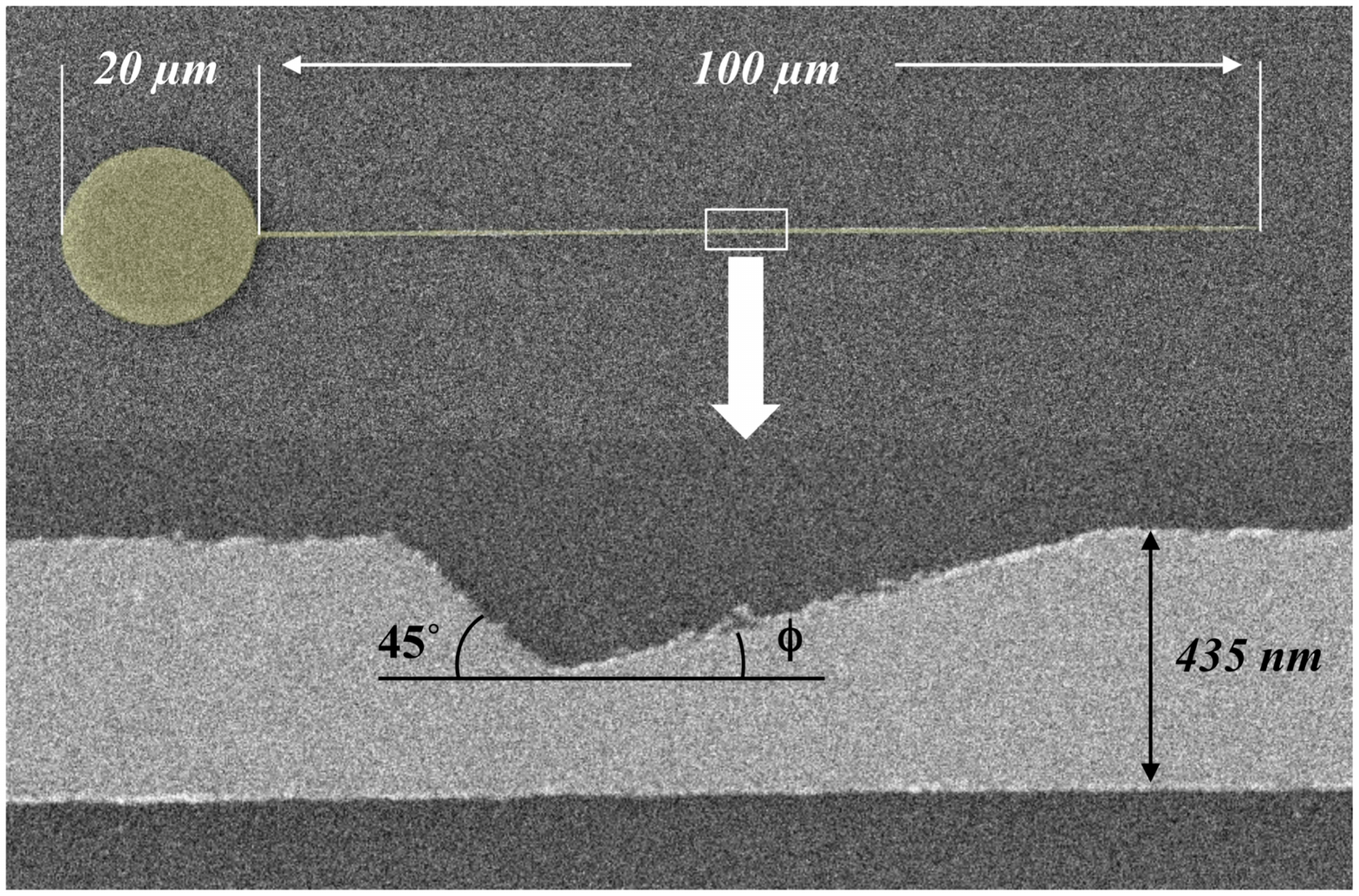}	
\caption{SEM image of a representative nanowire (top). Zoom of the notch placed in the middle of the wire (bottom). The angle $\phi$ defines the notch asymmetry (here, $15^o$). The red disks show the positions where the hysteresis loops were measured, before and beyond the notch. The vortex wall comes always from the pad. }
\label{Fig1}
\end{figure}

As can be seen from the scanning electron microscopy (SEM) image in Fig.~\ref{Fig1} the wires have three main parts: a pad 20 $\mu$m large in one end, a tapered shape in the other end, and at the middle, a triangular notch with a variable outgoing angle ($\phi$). Due to its smaller coercivity, domain walls are nucleated initially in the pad in both branches of the hysteresis cycle, i.e., for positive or negative magnetic fields. In each branch a different domain wall type will be nucleated, namely, HH for positive branch or TT on the negative one. Increasing the field up to a specific value (injection field), a wall is then injected into the wire as a vortex with CW or CCW chiralities and with a given polarity, positive ($p$) or negative ($n$). Due to the symmetry of the problem these four combinations of chirality and polarity take place irrespective of the wall type initially nucleated in the pad. We are assuming that all these eight different configurations are equiprobable when injected into the nanowire. After the injection, without any increment of the field, the wall is freely propagated through the nanowire until it starts to interact with the notch potential. Then, with further increase of the field, the wall will still be pinned until the field value is high enough to overcome the potential created by the notch. This is the depinning or coercive field. Again, without any further increment of the field, the wall will travel along the wire until be annihilated at the tapered end. 

The notches have a triangular shape with incoming angle kept constant at $45^o$ and their asymmetry being defined by the outgoing angle ($\phi$), which varies from $15^o$ to $45^o$. In order to guarantee good statistics and avoid any systematic error, arrays of 10 wires were grown for each $45^o-\phi$ symmetry, and all the them were measured. The distance between the notch vertex and the bottom wire edge was kept constant (235 nm) for all nanowires. 

To study the role played by the notch potential in the vortex wall dynamics, we have measured the coercivity along the wire through the focused MOKE. Our setup uses the MOKE in the longitudinal configuration with a focused 5 $\mu$m spot size. We have used a stable red laser diode ($\lambda=$ 635 nm) and a photodiode detector in differential mode. The applied magnetic field was continuously swept in a triangular waveform with frequency of $\approx$~1 Hz. The lock-in output signal feeds the oscilloscope, and hundreds of cycles are averaged after a couple of minutes. Single-shot measurements were also performed, though with a smaller signal-to-noise ratio. The sample holder is driven by motors with a minimum step of 28 nm in the $xy$ directions, which allows to measure hysteresis cycles scanning the wire. The setup is able to measure hysteresis cycles of single Py nanowires as small as 50 nm wide and 10 nm thick (further details will be published elsewhere). 

\section {Experimental results}

\begin{figure}[h] 
\includegraphics[width=0.95\columnwidth]{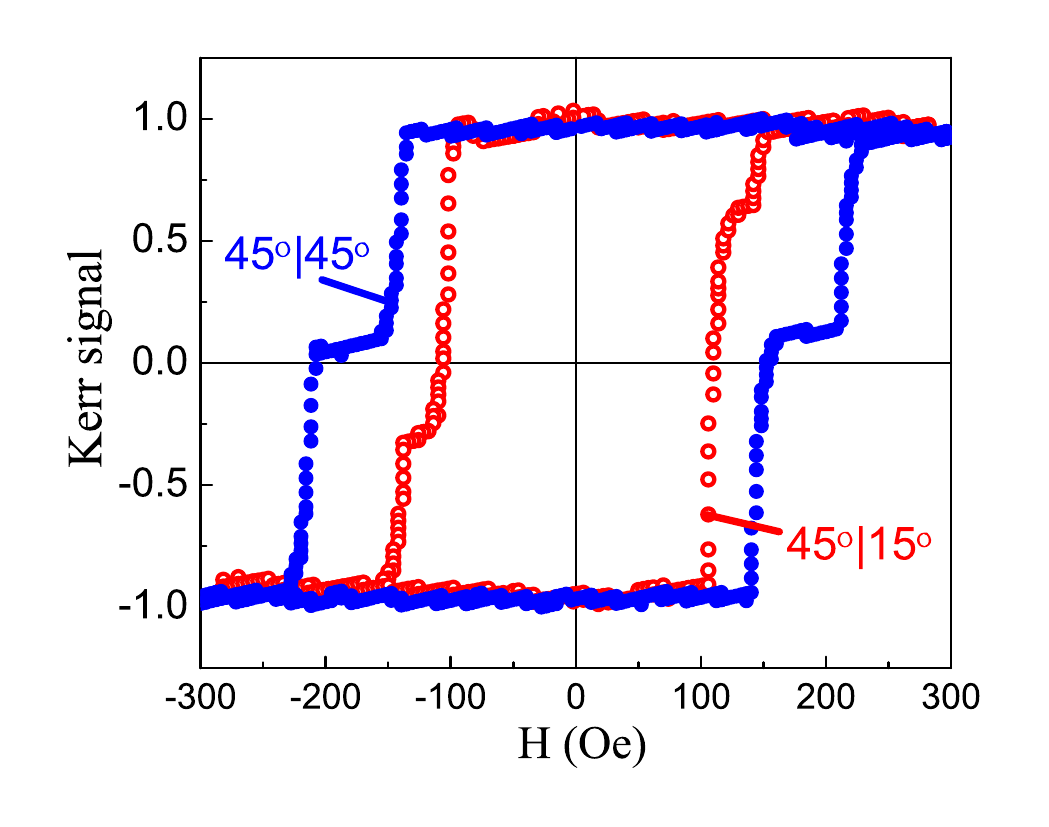}	
\caption{Averaged magnetic hysteresis cycles for a single wire measured beyond the notch for $45^o - 15^o$ (open red circles), and $45^o - 45^o$ (full blue circles).}
\label{Fig2}
\end{figure}

We firstly present averaged magnetic hysteresis measurements for single wires for $45^o-\phi$, with $\phi$ = $15^o$ and $45^o$ (see Fig.~\ref{Fig2}). The first is the most asymmetric notch we have prepared. The magnetic field is applied parallel to the wire. Measurements performed along the wire are quite similar when obtained at any point between the pad and the notch, or between the notch and the tapered end. The curves shown in Fig.~\ref{Fig2} were measured beyond the notch, at the middle section on the right of the notch. Before the notch the coercivity (37 Oe) is independent of the notch symmetry, as expected, and it is the same for all wires. It corresponds to the field necessary to inject a vortex wall from the pad into the wire and drag it up to the notch potential region. It is, then, the injection field. Beyond the notch the hysteresis cycles exhibit two coercivities, and it occurs for all values of $\phi$. In fact, all the 10 wires for each $45^o-\phi$ symmetry show similar cycles. It means that only the notch is being probed, eliminating the influence from possible defects at the wire edges. 

\begin{figure}[h] 
\includegraphics[width=1.0\columnwidth]{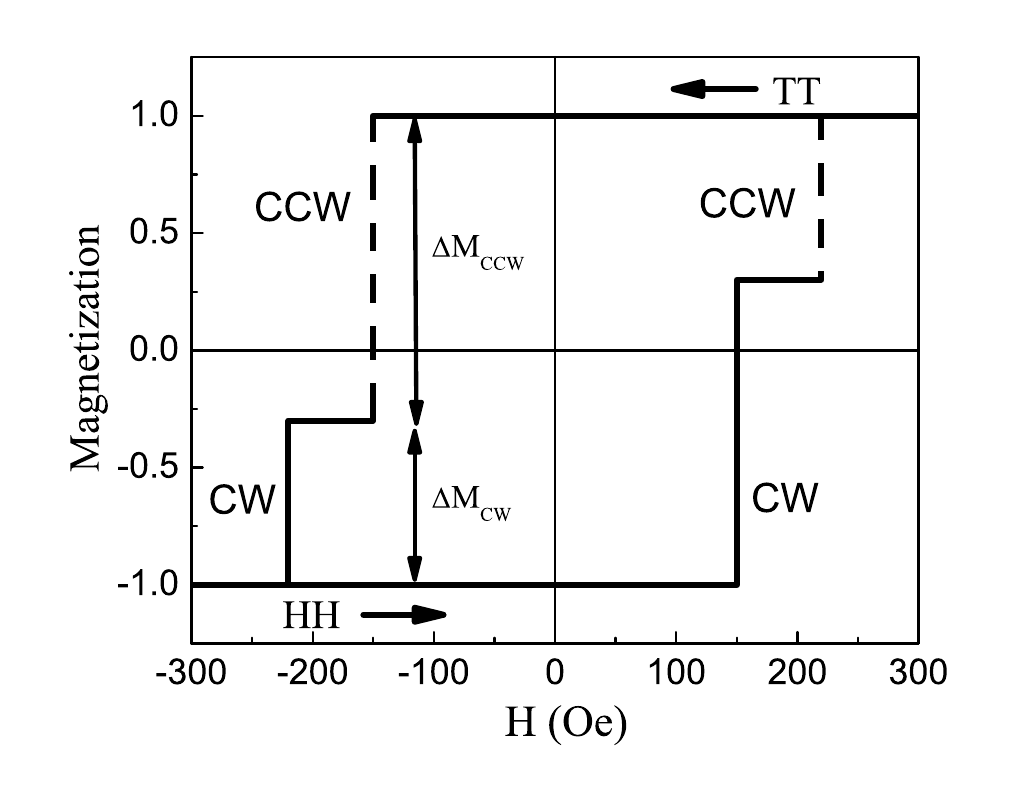}	
\caption{Sketch of an averaged magnetic hysteresis cycle for CCW and CW vortex walls in TT and HH configurations, respectively, reproducing hysteresis cycles measured by average (see text).}
\label{Fig3}
\end{figure}

As it is known, CW and CCW vortex walls in TT and HH configurations exhibit different coercivities when pinned by a notch~\cite{Bogart,Jin}. The depinning field of CCW (CW) chirality is smaller than the opposite chirality in the TT (HH) wall. It is sketched in Fig.~\ref{Fig3}. We follow our experimental procedure, adding the contribution of the CW and CCW chiralities that result from the average measurement protocol. Starting from the positive magnetic saturation, the vortex wall is nucleated as TT and stops at the notch. Increasing the field (in the negative sense) the wall is then depinned and the magnetization reversal takes place. The CCW wall is depinned for a field smaller than the CW one. Starting from the negative saturation, the opposite occurs -- the wall is nucleated as HH and when depinned, it is the CW chirality that exhibits smaller coercivity. This means that the hysteresis for a given chirality (CW or CCW) is shifted from zero field. This picture was also obtained from our micromagnetic simulations; we will return to this point later. 

\begin{figure}[h] 
\includegraphics[width=0.9\columnwidth]{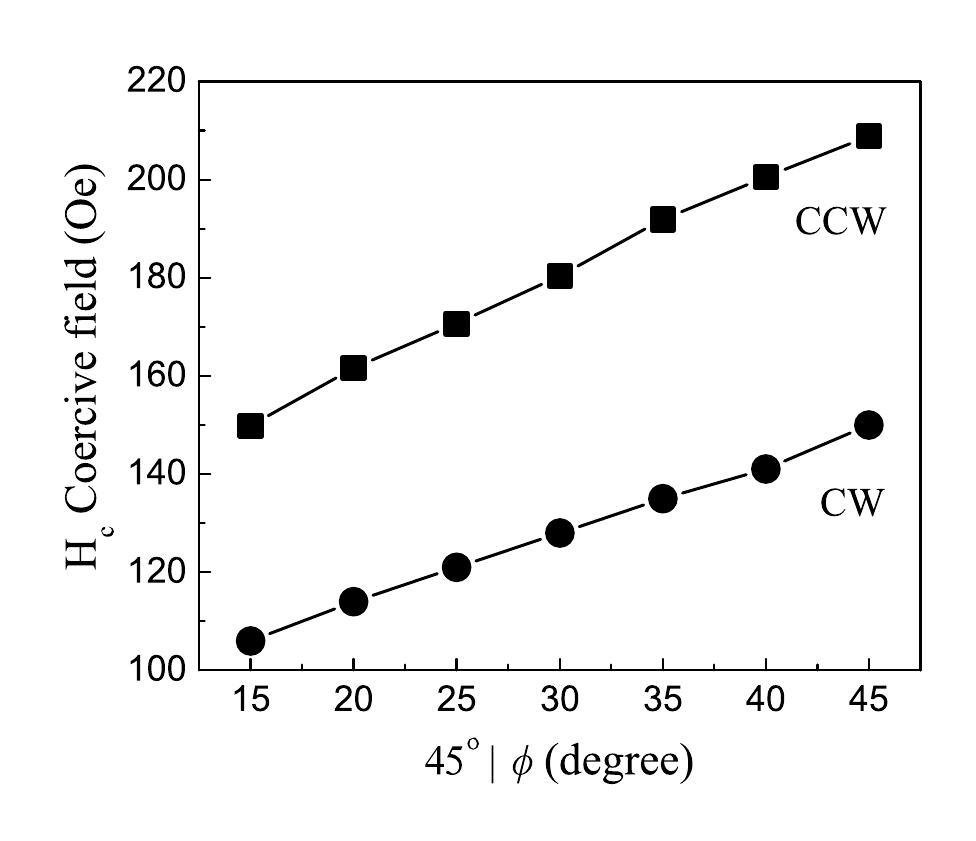}	
\caption{CW and CCW coercivity measurements changing the notch angle $\phi$. The walls are head-to-head. The lines are the linear best fit.}
\label{Fig4}
\end{figure}

Upon a closer look at the hysteresis cycles, a question about the magnetization value of the plateau between the two jumps comes up: if it is zero, it means that the two hysteresis curves have the same magnitude and then the CW and CCW chiralities are equiprobable, as expected. This is roughly true for $45^o-45^o$ (see Fig.~\ref{Fig2}). From Fig.~\ref{Fig3}, it should be expressed by the jumps $\Delta M_{CCW}$ and $\Delta M_{CW}$ with equal amplitudes. On the other hand, this is not absolutely true for the asymmetric notch, as can be verified in Fig.~\ref{Fig2}, for $45^o-15^o$. In the last, the amplitude of the magnetization jump of CCW depinning field is $\approx$~75$\%$ of the whole magnetization reversal, and this value is $\phi$ dependent. 

Fig.~\ref{Fig4} shows the $\phi$ dependence of the CW and CCW depinning fields. The wall is HH. The question of whether the asymmetry plays a role on the interaction between the vortex wall and the notch is also observed from the depinning field -- the strength of the potential energy produced by the notch varies linearly on $\phi$ and is maximum for the symmetrical shape. The depinning field is fitted by H$_C = 85.0 + 1.4~\phi$ for CW, and H$_C = 121.4 + 2.0~\phi$ for CCW. This dependence can be useful to tune the coercivity for practical applications. Actually, as we will show, other consequences from the asymmetry will emerge, e.g., on the dynamics of the vortex wall passing through the notch.  

\begin{figure}[h] 
\includegraphics[width=0.95\columnwidth]{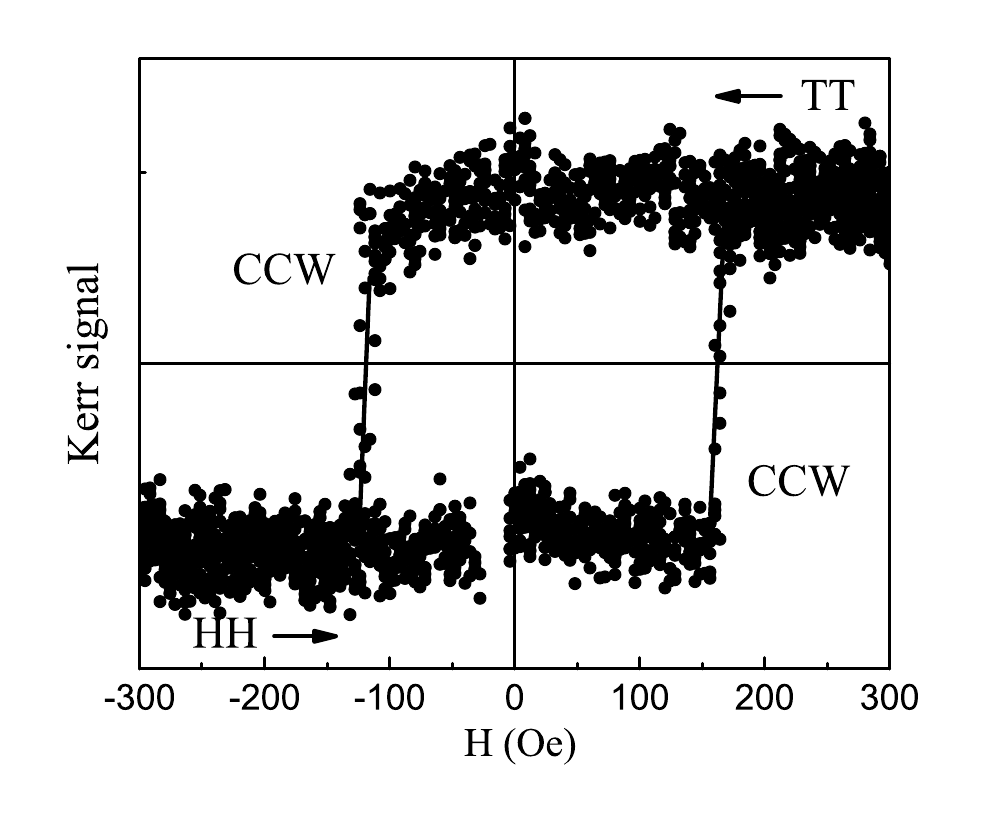}	
\caption{Single-shot hysteresis measurement for a single wire performed beyond the notch for $45^o-15^o$. The chirality is CCW at the magnetization reversal in negative (TT) and positive (HH) fields.}
\label{Fig5}
\end{figure}

The shift of the hysteresis cycles can be clearly seen from single-shot measurements performed for a single wire (see Fig.~\ref{Fig5}). In single-shot measurements any chirality can be observed in the magnetization reversal in negative (TT) and positive (HH) fields. To illustrate it, Fig.~\ref{Fig5} shows CCW walls in both negative and positive fields. The hysteresis is shifted +43 Oe from zero field. 

\begin{figure}[h] 
\includegraphics[width=1.0\columnwidth]{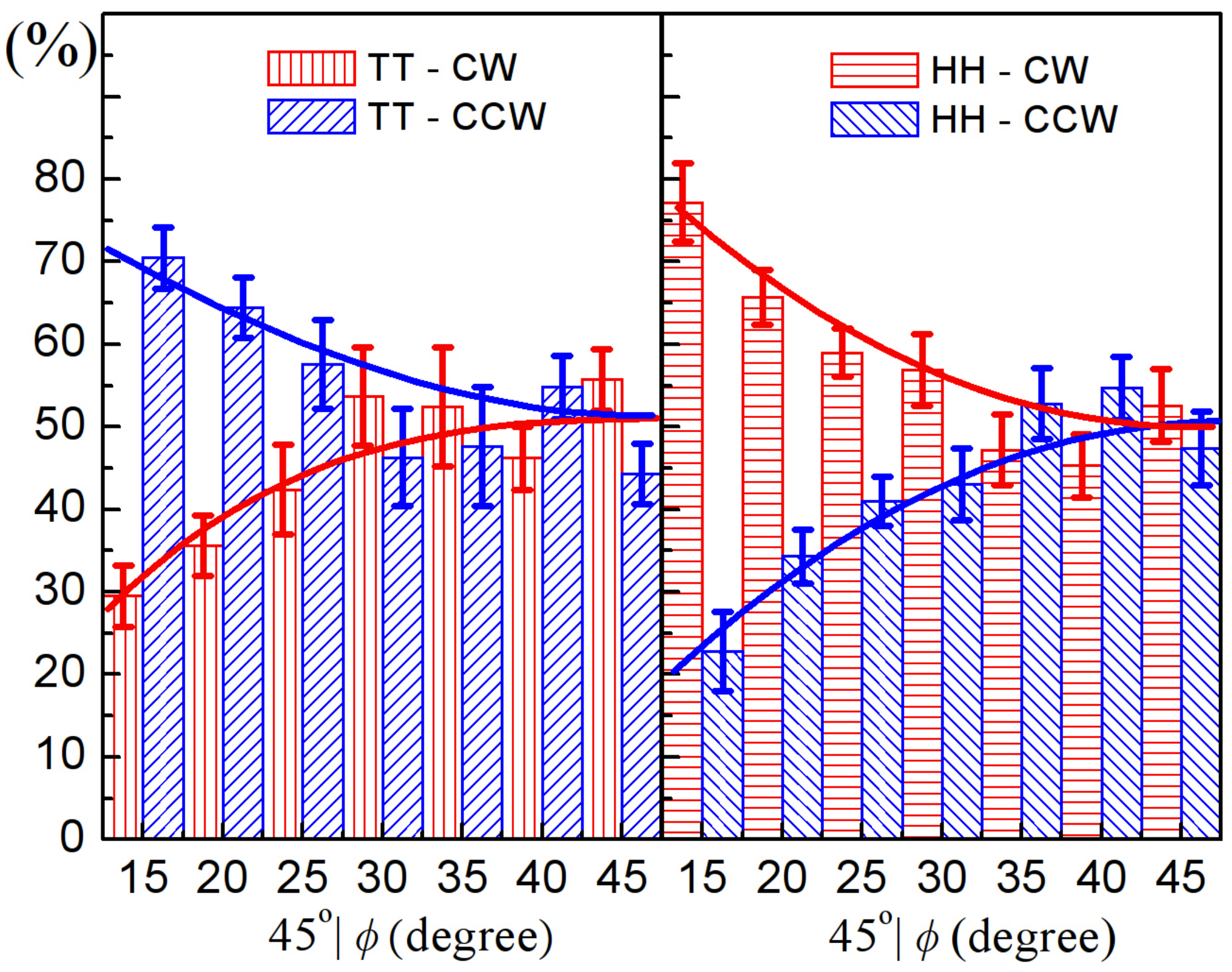}	
\caption{CW and CCW occurrence probability histogram obtained from hysteresis cycles measured for TT and HH walls for the notch with $45^o - \phi$ symmetry, with $\phi$ varying from $15^o$ to $45^o$. The lines are guides for the eyes.} 
\label{Fig6}
\end{figure}

To investigate the relative occurrence of CW and CCW walls we have performed hysteresis measurements like those shown in Fig.~\ref{Fig2}, i.e., doing averages with the oscilloscope. The occurrence probability of CCW (CW) walls was obtained by $\Delta M_{CCW (CW)} / (\Delta M_{CCW} + \Delta M_{CW})$. We performed it for all the 10 wires to eliminate any possible influence from defects at the wire edges that could produce an extra pinning of the wall. We have thus obtained the occurrence probabilities for TT and HH configurations, varying $\phi$ from $15^o$ to $45^o$ (see Fig.~\ref{Fig6}). The dispersion around the average varies from 4\% to 7\%. To be sure that the probabilities are not influenced by some field misalignment, we repeated the measurements for one of the wires changing the angle between the field (in the plane of substrate) and the wire from $+5^o$ to $-5^o$ and no significant change was observed. We also obtained the occurrence probabilities from single-shot measurements, instead of averaging the hysteresis with the oscilloscope. Repeating single-shot measurements 10 times for each one of the 10 wires and then calculating the average, one obtained approximately the same probabilities shown in Fig.~\ref{Fig6}, but with larger dispersion (15\%). 

We emphasize that we took several precautions on the CW and CCW occurrence probability measurements, mainly because the results are unexpected. As can be seen from Fig.~\ref{Fig6} for small $\phi$, the CCW wall is more likely to occur than the CW wall on the wire magnetization reversal at negative field, i.e., for TT walls. Whereas for positive field, HH walls, the CW wall is more probable. In any case the imbalance of the probabilities is close to a factor of 2.5 -- 3. 

\section{Micromagnetic Simulations and Discussion}

In order to investigate the magnetization reversal process in further detail, mainly including the chirality flipping, we have performed micromagnetic simulations. We have used OOMMF \cite{oommf} and MuMax3 \cite{mumax} codes with typical parameters for Py: $A = 13\times10^{-12}$ J/m, $M_{s} = 860\times10^{3}$ A/m, and $\alpha = 0.005$ \cite{tico} in the dynamics analysis, where $A$ is the exchange constant, $M_{s}$ the saturation magnetization, and $\alpha$ is the damping constant. In all simulations the wires are 10 $\mu$m long, $400$ nm wide and $30$ nm thick, and are discretized in cells of $5\times 5\times 30$ nm$^3$. The center of the notch is placed 5 $\mu$m from the left end and it has the same geometry as found in the experiments, i.e., the incoming angle was kept with $45^o$ and the outgoing angle $\phi$ varies from $15^o$ to $45^o$. To simulate the hysteresis cycles, though, we have used $\alpha$ larger than those experimentally accepted to have the convergence criterium quickly achieved, so we used $\alpha = 0.5$. 

\subsection{Simulation of the Vortex Wall movement}

\begin{figure}[h] 
\includegraphics[width=0.9\columnwidth]{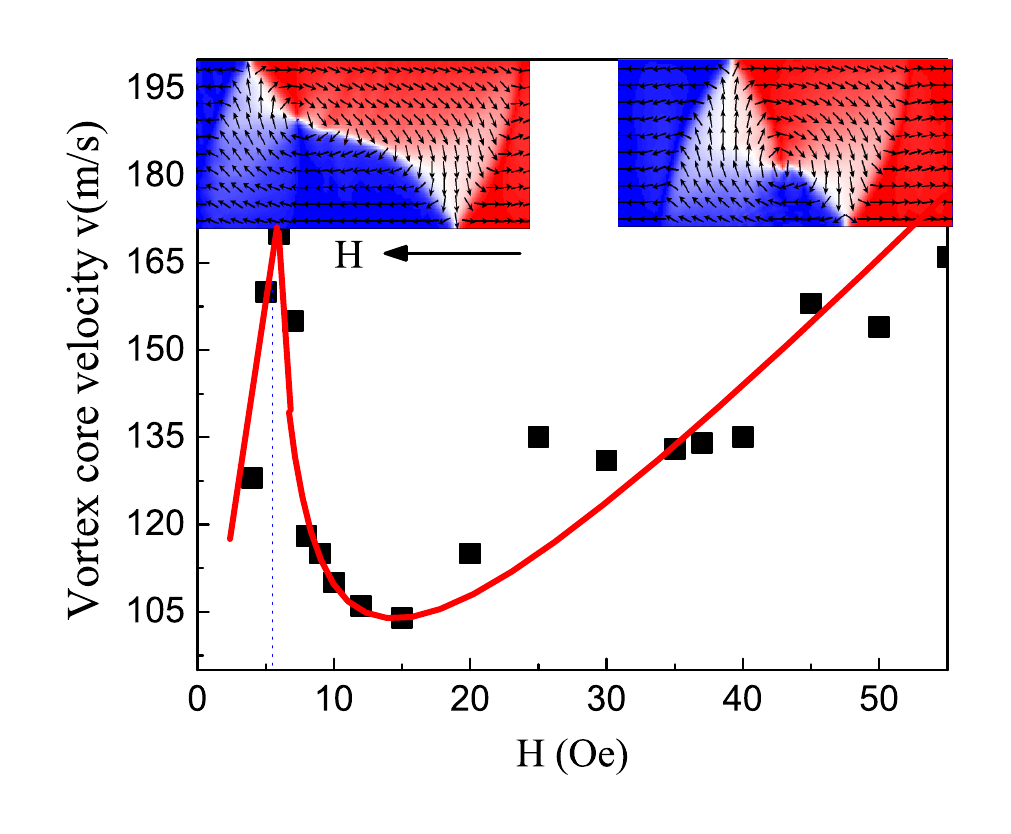}	
\caption{Simulated vortex wall velocity as a function of the applied magnetic field. The line is a guide for the eyes. Inset: Snapshots of the vortex wall moving to the right of the Py wire under a field of 40 Oe. The vortex core moves along the line that separate the domains and the wall width oscillates.}
\label{Fig7}
\end{figure}

As it is well known, vortex walls propagate in thin wires (or flat stripes) under action of magnetic field, and depending on the field intensity the wall propagation follows different regimes. Above a critical field the wall velocity is no longer linear on field but it drops, a phenomenon known as the Walker breakdown, and this critical field intensity is the Walker field~\cite{Atkinson,Yang,Walker,Guimaraes}. The core movement of the vortex wall along the wire changes from steady-state to precessional regime, where besides the drift movement the core position oscillates between the bottom and top edges of the wire~\cite{Clarke,Thiaville,Tchernyshyov}. Experiments on the vortex wall velocity in Py wires have reported the Walker field as close to 5-10 Oe~\cite{Erskine,Stuart}. This critical value is well below our measured injection field (37 Oe), indicating that the vortex wall dynamics in our experiments is above the Walker field. We have simulated the vortex wall velocity as a function of the magnetic field (see Fig.~\ref{Fig7}). Notice that there is no notch in these simulations. The simulated Walker field found is close to 5 Oe, which is in agreement to values found in experiments for Py wires~\cite{Erskine,Stuart}. In addition, the wall velocities ($\approx$~100-150 m/s) are also in agreement with experiments~\cite{Erskine,Masamitsu}. 

In order to know in further detail the wall movement in the precessional regime, we have simulated the vortex wall propagation for a field of 40 Oe, that is approximately the experimental injection field. Snapshots of the CW vortex wall movement along the wire in TT configuration are shown in the inset of Fig.~\ref{Fig7}. Notice that only a partial view of the wire is shown, the whole wire was taken into account in all simulations. Besides the wall movement to the right, the vortex core moves along the line that separate the domains. It goes back-and-forth reversing the core polarity close to the edges, when the polarity is negative the core moves upward along the line, when the polarity is positive it moves downward. This movement is driven by the gyrotropic field which is given by pG$\widehat{z} \times \overrightarrow{v}$, where p is the polarity ($\pm1$), and $\overrightarrow{v}$ the core velocity \cite{Thiele,Clarke,Thiaville2}. G is the gyrovector, G$=2\pi \mu_0 M_s d/\gamma$, where $\mu_0$ is the vacuum permeability, $M_s$ is the magnetic saturation, d is the thickness, and $\gamma$ is the gyromagnetic ratio. 

In addition to the oscillatory vortex core movement the wall width also changes - the core moves upward and the wall contracts, when the core moves downward the wall expands. Defining the wall width as the distance between the two points of the diagonal line that separate the domains, the wall width oscillation amplitude is 240 nm, approximately, varying from 420 to 660 nm. After injection into the wire, this is the dynamics of the vortex wall traveling along the wire toward the notch. As mentioned before, the vortex wall moves, stops at the notch until the field reaches the depinning field (100-200 Oe, see Fig.~\ref{Fig4}) to pass through the notch, completing the magnetization reversal process.  

\subsection{Simulation of the Vortex Chirality flipping}

\begin{figure}[h] 
\includegraphics[width=0.55\columnwidth]{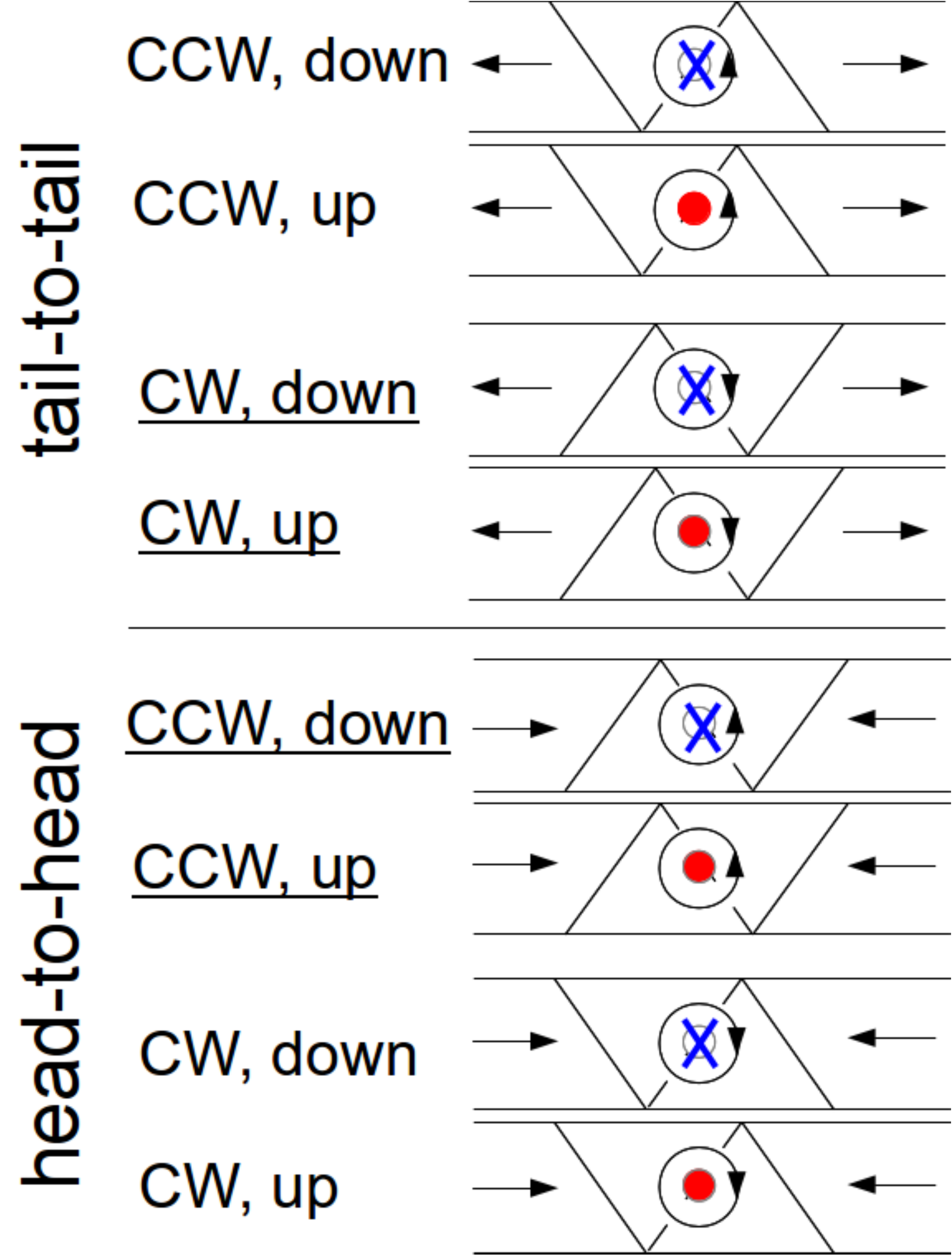}	
\caption{Sketch of the vortex wall chirality in TT and HH configurations. The core polarity is indicated (negative, blue cross and positive, red dot). The vortex wall moves to the right side. The configurations which change the chirality are underlined. The vortex core polarity does not play any role on the chirality flipping process.} 
\label{Fig8}
\end{figure}

Let us now examine what happens to the vortex wall when it reaches the notch. According to Fig.~\ref{Fig6}, for some events in TT configuration, the CW wall is transformed to CCW, while for HH it is the opposite, the CCW wall is transformed to CW. Fig.~\ref{Fig8} shows a sketch of the vortex wall shape according to the chirality and polarity. It shows the symmetry between the TT and HH configurations. The vortex wall is under applied field and is moving to the right. Depending on the chirality and wall configuration, the line at the right side of the vortex wall leans to the left or to the right. For TT, the CW wall has the right side leaned to the right and when it arrives at the notch, this line is approximately perpendicular to the notch, the $45^o$ side. On the other hand, for the HH this line is approximately parallel to the edge of the notch. It is reasonable to assume that only in the first case any changing or vortex chirality flipping can be expected because the wall structure is disturbed. 

\begin{figure}[h] 
\includegraphics[width=0.7\columnwidth]{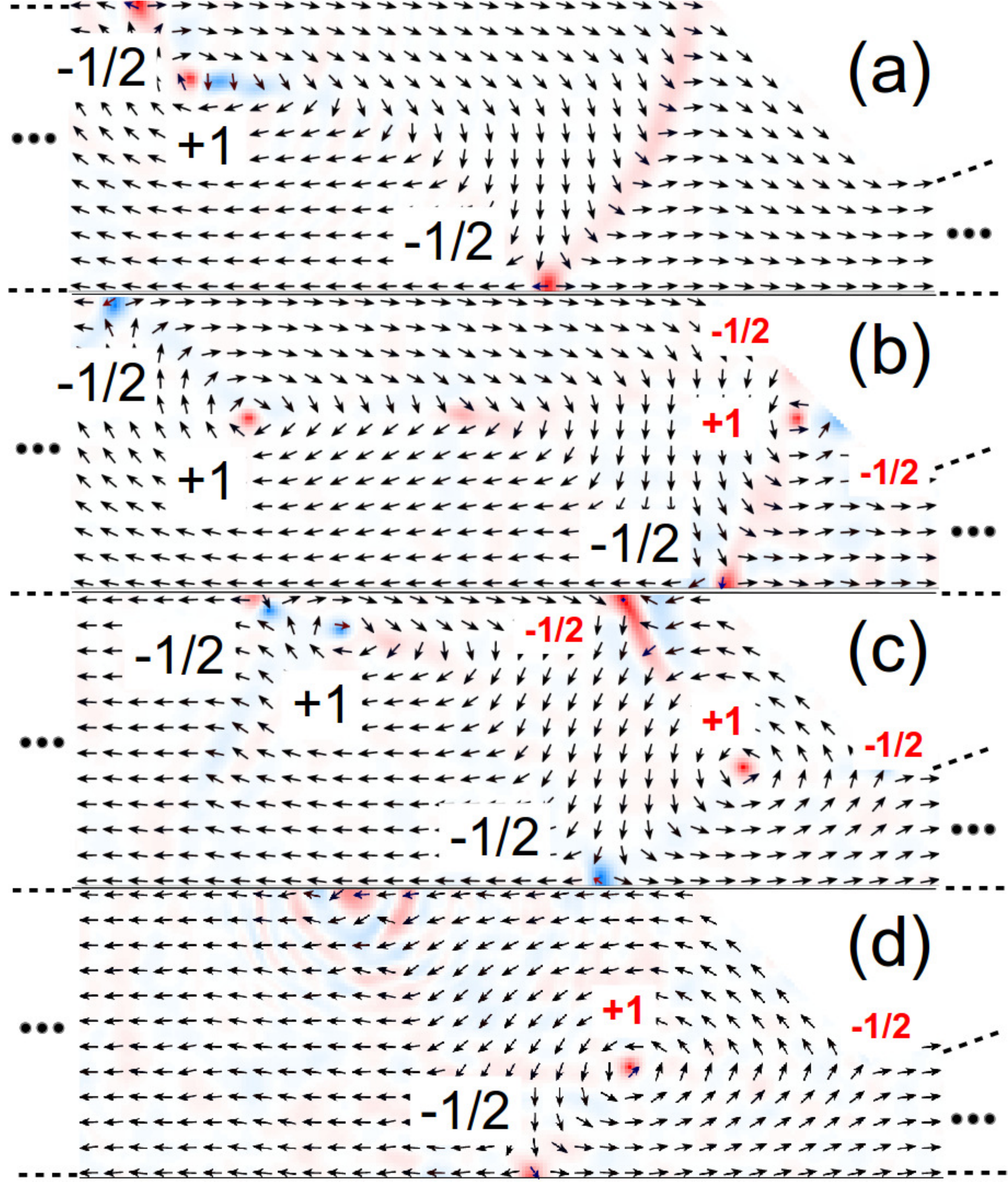}	
\caption{Snapshots of the magnetization dynamics with an incoming CW vortex wall in the TT configuration. The notch is $45^o-15^o$. (a), (b), (c), and (d) are subsequent instants. The chirality flipping from CW to CCW is observed; +1 and -1/2 are winding numbers (see text).} 
\label{Fig9}
\end{figure}

Let us see what happens to the configurations where the chirality flipping takes place. Some snapshots of the the chirality reversal dynamics for $45^o-15^o$ from CW to CCW for the TT configuration are shown in Fig.~\ref{Fig9}. The simulations were performed taking into account the whole wire, and not only the area around the notch. The  vortex wall moves toward the notch and at a given moment the line on the right of the vortex wall touches the notch. As can be seen from Fig.~\ref{Fig9}b, a new vortex is nucleated at the notch edge with CCW chirality, and the original vortex reverses its polarity and goes up to the upper wire edge (Fig.~\ref{Fig9}c). Subsequently, it disappears remaining only the CCW vortex wall (Fig.~\ref{Fig9}d). No dependence of the vortex core polarity on the chirality reversal process was observed. 

At this point it is useful to use concepts of topological defects to go further on the chirality flipping dynamics~\cite{Mermin}. Topological defects are given by winding numbers, defined by $m_b = 1/2\pi \oint \nabla\theta dr$ if they are in the bulk or $m_e = 1/2\pi \oint \nabla(\theta-\theta_\tau) dr$ if the defects are at the edges of the magnetic structure~\cite{Chern,Pushp,Tchernyshyov}. For the bulk the integration involves the defect and it is $\pm1$. For the edges it is a fractional number, $\theta_\tau$ is an angle taking the tangent directions of the edge. Vortex walls have one bulk defect in their center with $m_b=+1$ and two edge defects with $m_e=-1/2$, one in the upper edge and the other in the bottom edge of the wire. For vortex walls the total winding number is $m=m_b+m_e=0$, and this number is conserved along any movement or dynamics of the vortex wall~\cite{Tchernyshyov,Pushp}. 

Fig.~\ref{Fig9} shows the winding number for the bulk and edge topological defects for the original (large numbers) and nucleated vortices (small numbers). As can be seen, the nucleated vortex has one of the edge defects that takes different dynamics. The new upper edge defect moves along the wire edge to be annihilated with the original upper edge and bulk defects, remaining the original bottom edge with the nucleated bulk and edge defects. The new nucleated bulk defect has the reversed chirality. We found this process in simulations varying the $\phi$ angle from $15^o$ to $40^o$. However, for the symmetric $45^o-45^o$ notch the chirality flipping was not observed.  

Still from simulations, we have observed that the vortex wall is at its maximum width when the chirality flipping takes place. Actually, the chirality flipping process begins when the wall contracts from the maximum size which leads to the nucleation of the extra vortex. The reason for this nucleation is not fully understood, though the simulations  are useful to describe the general flipping process. 

As mentioned before, the vortex wall moves with oscillations on the width and the core position that goes back-and-forth between the wire edges. Besides this, one observed that the core movement is not fully regular, it is interrupted by periods moving at the upper and bottom edges. The duration for such periods is irregular. Therefore, depending on the position where the vortex wall starts its movement the core has a different path and the wall arrives at the notch with different wall width. Depending on the starting position we observed that sometimes the chirality flipping does not take place. When it occurs, the wall width is at its minimum size. For other intermediate wall widths, the vortex at the notch edge is not nucleated, or it is, but it lasts for a short time and is annihilated. Thus, whether or not the chirality flipping takes place it depends on the wall state, i.e., if it arrives at the notch contracting or expanding.

\subsection{Simulation of the Vortex Chirality Occurrence Probability}

In the experiments we measured beyond the notch different occurrence probabilities for the CW and CCW chiralities, e.g., see Figs.~\ref{Fig2} and \ref{Fig6}. The probability measured in experiments is related to stochastic processes provided by the temperature, to eventual defects at the wire edges, and as suggested by simulations, to a non regular vortex wall movement. In our simulations the last one could easily be included, just varying the initial position of the vortex wall. 

\begin{figure}[h] 
\includegraphics[width=0.85\columnwidth]{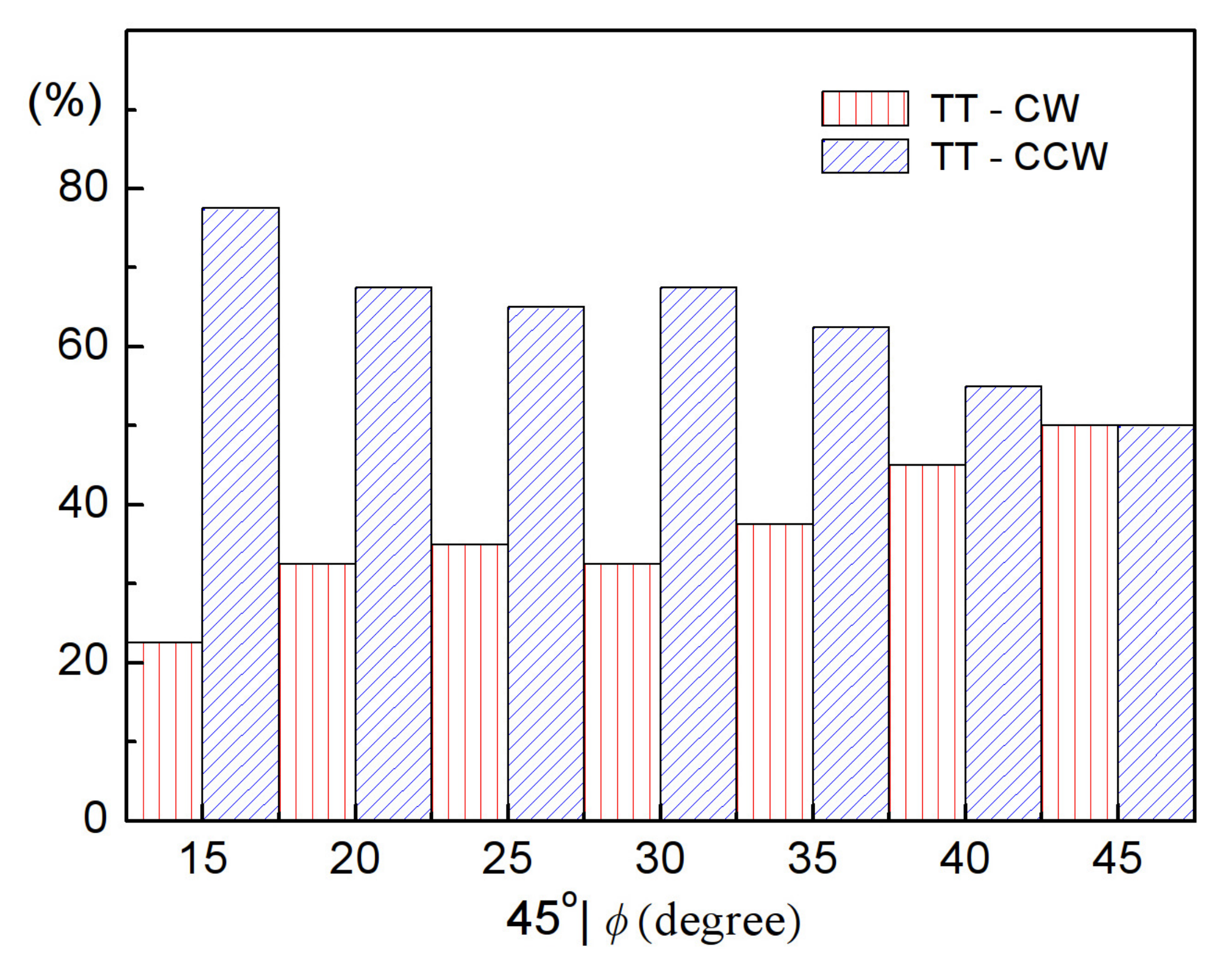}	
\caption{Simulated occurrence probability of CW and CCW walls as a function of $\phi$. The wall configuration is tail-to-tail. A fraction of CW walls after to interact with the notch flips its chirality to CCW.}
\label{Fig10}
\end{figure}

We have simulated the vortex wall dynamics starting the wall movement from positions ranging from 0.8 to 3.4 $\mu$m away from the notch. The walls begin as CW and CCW in the TT configuration. Concerning the core polarity, we have used positive and negative polarities with the same probabilities. We simulated the vortex wall dynamics from the starting position to the position it interacts with the notch, leading or not to a changing on the chirality, and then stopping. After that, increasing the field the vortex wall pass through the notch keeping the chirality. It reproduces the experimental protocol. Except by the starting position, all the parameters used in the simulations are the same. No dependence of the core polarity on the chirality flipping was found.   

Varying the angle $\phi$ we simulated the dynamics for 20 values of the position and calculated the relative number of chirality reversal events. For CCW walls, irrespective of $\phi$, no event with chirality flipping was observed. On the other hand, for CW walls in a fraction of the simulated events the transformation from CW to CCW occurred, and it is dependent on $\phi$. 

Fig.~\ref{Fig10} shows the dependence on $\phi$ of the relative number of CW and CCW walls at the notch. For $\phi$~=~$15^o$ one observes that $\approx$~55\% of the CW changed to CCW, and as the CCW walls do not switch, so we have $\approx$~75\% ((0.55+1)/2) of CCW walls. For the other $\phi$ range side, the symmetric $45^o-45^o$ notch, no change from CW to CCW was observed, meaning that the probability to find the CW or CCW walls is 50\%. Between $15^o$ and $45^o$, from the observed linear variation, one can clearly see that the notch symmetry plays a role on the chirality flipping process. Our simulations show a good agreement with the experimental results.  

\section{Conclusions}

Summarizing, we have shown that Py nanowires with triangular asymmetric notches can be used to manipulate vortex walls in narrow wires, including its dynamics in the magnetization reversal. Through the focused longitudinal MOKE  we measured hysteresis curves before and beyond the notch with a good signal-to-noise ratio, which allowed  to probe clearly the dependence of the chirality on the magnetization reversal dynamics. The coercivity is larger for CCW than CW for HH walls, and the opposite is valid for TT, and it increases linearly on $\phi$ reaching a maximum at $45^o-45^o$. 
 
This chirality selectivity and this linear variation of the depinning field can be useful for practical applications. Notches with high asymmetry, e.g., $45^o-15^o$, act like a chirality filter. According to the wall type, if TT or HH, the CCW or CW wall prevails, respectively, amounting to a high percentage $\approx75\%$. Simulations based on the Landau-Lifshitz-Gilbert equation were able to describe the main characteristics of the chirality flipping process and to obtain the relative number of flipping events in good agreement with the experimental results.

\section{Acknowledgements}

We acknowledge financial support from the Brazilian agencies CNPq, CAPES, FAPERJ, and INCT-INFO. We also acknowledge the LABNANO/CBPF for the use of the nanolithography facilities and the CAT/CBPF for the computer facilities. We thank A.P. Guimar\~{a}es for helpful discussions.


\begin{thebibliography}{10}
\bibitem {tico} For Py films grown in the same sputtering with the same growth conditions as the nanowires, we have measured $\alpha$ between 0.005-0.008 through a broad-band ferromagnetic ressonance setup.
\end{thebibliography}
\end{document}